\documentclass{conm-p-l}
\usepackage{amsthm,amsfonts,latexsym,amssymb
}
\usepackage[dvips]{graphics}

\def\a{\alpha}
\def\b{\beta}
\def\c{\gamma}
\def\g{\gamma}
\def\d{\delta}
\def\e{\varepsilon}
\def\s{\sigma}

\def\D{\Delta}

\def\l{\lambda}
\def\L{\Lambda}
\def\p{\partial}

\def\R{\mathbb R}
\newcommand{\ket}[1]{\vert{#1}\rangle}
\newcommand{\bra}[1]{\langle{#1}\vert}
\def\R{\mathbb R}
\def\Z{\mathbb Z}

\newcommand{\be}{\begin{equation}}
\newcommand{\ee}{\end{equation}}
\newcommand{\bea}{\begin{eqnarray}}
\newcommand{\eea}{\end{eqnarray}}
\newcommand{\beax}{\begin{eqnarray*}}
\newcommand{\eeax}{\end{eqnarray*}}

\newcommand{\mfr}[2]{{\textstyle\frac{#1}{#2}}}

\newcommand{\lcm}{\mbox{\rm lcm}}
\newcommand{\cl}{\mbox{\rm cl}}

\newtheorem{theorem}{Theorem}[section]
\newtheorem{lemma}[theorem]{Lemma}
\newtheorem{hypo}[theorem]{Hypothesis}
\newtheorem{assumption}[theorem]{Assumption}

\theoremstyle{definition}

\begin{document}

\title[Dynamics of interfaces in the ferromagnetic XXZ chain]{On the dynamics of 
interfaces in the ferromagnetic {XXZ} chain under weak perturbations}

\author{Bruno Nachtergaele}
\address{Department of Mathematics, University of California, Davis, One Shields 
Avenue, Davis 95616-8366, USA}
\email{bxn@math.ucdavis.edu}
\thanks{This material is based on work supported by the National Science
Foundation under Grant No.~DMS0070774. Shannon Starr is a National Science 
Foundation Postdoctoral Fellow}
\copyrightinfo{2002}
    {by the authors. This article may be
     reproduced in its entirety for non-commercial purposes.}
\author{Wolfgang L Spitzer}
\address{Department of Mathematics, University of California, Davis, One Shields 
Avenue, Davis 95616-8366, USA}
\email{spitzer@math.ucdavis.edu}
\author{Shannon Starr}
\address{Department of Physics, Princeton University,
Princeton, NJ 08544, USA}
\email{sstarr@math.princeton.edu}
\date{September 30, 2002}


\begin{abstract} We study the time evolution of interfaces of the 
ferromagnetic {XXZ} chain in a magnetic field. A scaling limit is 
introduced where the strength of the magnetic field
tends to zero and the microscopic time to infinity while keeping 
their product constant. The leading term and its first correction 
are determined and further analyzed in more detail for the
case of a uniform magnetic field. 
\end{abstract}

\maketitle

\section{Introduction}

The quantum ferromagnetic Heisenberg model has received a great deal of
interest in recent years.  We shall concentrate here on the one-dimensional
model. The starting point was the discovery of interface ground states in the
anisotropic {XXZ} Heisenberg chain by Alcaraz, Salinas and
Wreszinski~\cite{ASW} and independently by Gottstein and Werner~\cite{GW}.
These interface ground states display  two regions of opposite magnetization
separated by a domain wall whose thickness depends on the anisotropy. 
Explicit expressions for the ground states exist for all values of the spin, 
see Section~\ref{model}. 

The next important contribution was the proof of completeness of the 
infinite volume ground states (in the sense of local stability) and the 
proof of a positive gap above the ground states for any amount of anisotropy 
and any value of the spin. The spin-$\frac{1}{2}$ model is $SU_q(2)$ 
symmetric which was used to determine the exact value of the gap. This was 
accomplished by Koma and Nachtergaele~\cite{KN1,KN3} and 
Matsui~\cite{Mat}. 

The existence of a spectral gap for general $j$ was proven by Koma,
Nachtergaele and Starr~\cite{KNS}.  Recently, it was proven by Caputo and
Martinelli~\cite{CapMart2} that the gap for spin $j$ scales asymptotically 
like $j$, however, no  precise values are known. This verifies a conjecture 
stated in~\cite{KNS}, and  comes as a surprise since the spin-$j$ Ising chain 
(which can be considered as the XXZ chain with infinite anisotropy) always 
has a gap equal to 1. 

There has been a lot of progress over the last two decades on the dynamics of
classical spin systems, or lattics gas systems. For instance,  for the Ising
model the Glauber and Kawasaki dynamics have been studied extensively. Here, we
only want to mention the  work by Schonmann and Shlosman~\cite{SS} on the
relaxation in the Ising model which is  similar to what we want to study in the
quantum XXZ chain. 
A quantum model which has been studied in detail is the one-dimensional
XY model, by Antal, R\'acz, R\'akos, and Sch\"utz~\cite{ARRS} at zero
temperature and by Ogata~\cite{Oga} at non-zero temperature. More recently,
Aschbacher and  Pillet~\cite{AP} studied non-equilibrium steady states in the
same model at non-zero temperature. This is of great interest also in the
context of entropy production. The XY model (in one dimension) is easier to
deal with  than the XXZ model since it can be mapped to a model of free
fermions (see~\cite{LSM} and~\cite{Araki2}).  

Interfaces and their response to magnetic fields are of great physical  and
technological interest, for instance   in storing and reading information on
magnetic devices. The concept of a  magnetic domain was first introduced by
Pierre Weiss, already in 1906. It was in 1935 in a seminal paper that Landau
and Lifshitz presented a theoretical treatment of domain walls and their motion
when a magnetic field is applied. Although the Landau-Lifshitz equation was
rigorously derived in the mean-field limit by Moser, Prets, and Spitzer
~\cite{MPS}, its predictions do not universally hold true for all
magnetic materials. Thus we want to understand  the dynamical behavior of
domain walls in a quantum model, in which they occur naturally as ground
states, namely in the XXZ model. 

The dynamics we consider is generated by the {XXZ} kink Hamiltonian  perturbed
by a magnetic field. As in many similar cases, we introduce a scaling where the
strength of the perturbation, say the coupling constant $\lambda$ for a 
magnetic field, tends to 0 and the  microscopic time scale to infinity such
that their product, which we denote by $\tau$, is kept fixed. We
obtain the following results about this limit: 
(i) For external magnetic fields of bounded spatial support that are
sufficiently regular functions of time, the leading contribution to the time
evolution is identified as the reduced dynamics determined by the time averaged
field projected onto the ground state space of the unperturbed model with an
error term of order $\lambda^{1-\delta}, \delta \in (0,1)$ (Theorem
\ref{leading});
(ii) If the spectrum of the unperturbed model has a gap between the ground
states and the rest of the spectrum, and the rest of the spectrum is absolutely
continuous, the next-to-leading order term is of order $\lambda$ (Theorem
\ref{thm:correction});
(iii) Analysis of the reduced dynamics for uniform, time-independent magnetic 
fields, reveals a markedly different behavior depending on whether or not the 
field has a non-vanishing component in the $z$ direction. If the $z$ component
vanishes, the reduced dynamics for the magnetization
profile is ``ballistic'', while if there is a non-vanishing component of the
field in the $z$ direction, the magnetization profile evolves periodically.

In Section \ref{model} we define the model and state the assumptions on the 
magnetic field. We derive the leading dynamics and its first correction in 
Section \ref{dynamics}. In Section \ref{leading dynamics} we analyze the
leading dynamical behavior in more  detail for a uniform field and comment on
small perturbations thereof. Two appendices provide some auxiliary results
on the magnetization profile in the kink ground states of the XXZ chain
and on the spectrum of the Stark-Jacobi operator.

\section{The model}\label{model}

\subsection{Finite chain Hamiltonian}

In this paper we shall only consider the
spin-$\frac{1}{2}$ chain. Unless stated otherwise,
our results extend to higher values of the spin. 
As usual, let us denote by $\D>1$ the anisotropy
parameter. The kink Hamiltonian on the chain $[a,b]\cap\Z$ is defined as 
\be\label{H}
    H[a,b] = -\sum_{x=a}^{b-1} \left[\frac{1}{\D}(S^1_x S^1_{x+1}
    + S^2_x S^2_{x+1})+ S^3_x S^3_{x+1} - \frac{1}{4}{\bf 1}\right] -
    \frac{1}{2}\sqrt{1-\D^{-2}}
    (S_a^3 - S_b^3)\,  ,
\ee 
where the spin operators at position $x$,
$S_x^1,S_x^2,S_x^3$ satisfy the usual commutation relations, $[S_x^\a,S_x^\b]
= i\,\e^{\a \b \c}\,S_x^\c$. We have added the kink boundary condition,  
$-\frac{1}{2}\sqrt{1-\D^{-2}}
(S_a^3 - S_b^3)$ which makes the model $SU_q(2)$ invariant and displays the 
full set of kink ground
states which we derive next. It is convenient to introduce another parameter, 
$0<q<1$, such that $q+q^{-1}=2\D$. One can show that 
$$H = \sum_{x=a}^{b-1} P_{x x+1}^q \, ,
$$ 
where $P_{x x+1}^q$ is the orthogonal projection onto the vector 
$q\ket{\uparrow\downarrow}- \ket{\downarrow\uparrow}$. By introducing the 
$q$-deformed lowering operator
$$S^- = \sum_{x=a}^b {\bf 1}\otimes \cdots \otimes S_x^-\otimes t_{x+1}
\otimes\cdots\otimes t_b
$$
with $t=q^{2S^3}$, we can express all $b-a+2$ ground states as
$(S^-)^k\ket{\uparrow\cdots\uparrow}, k=0,1,\ldots,b-a+1$,
since $[H,S^-]=0$. Clearly, these vectors run through all
possible eigenspaces of the total $S^3$ component, which obviously commutes 
with $H[a,b]$. The uniqueness of the ground states follows, for instance, 
by a Perron-Frobenius argument.

More explicitly, let us denote the eigenvectors of the total $S^3$ component 
by $\ket{(m_x)}$, where $(m_x)\in \{\pm\frac{1}{2}\}^{b-a+1}$ and $S_y^3
\ket{(m_x)} = m_y \ket{(m_x)}$. 
Then, as was found by Alcaraz, Salinas and Wreszinski~\cite{ASW}, these 
ground states are (up to normalization)
\be
 \psi_m = \sum_{(m_x)} \prod_{x=a}^b q^{-x(1-m_x)/2} \ket{(m_x)} \, ,
\ee
where the sum runs over all sets $(m_x)$ such that $\sum_x m_x=m$, which 
is the total $S^3$ eigenvalue.
As the $\psi_m$ are not normalized, we also define $\ket{m}=\psi_m/\|
\psi_m\|$. If $m=\pm(b-a+1)$ is maximal/minimal, then we have the all 
spin-up/down state, i.e., the magnetization profile in the $z$ direction 
equals $\bra{m} S_x^3\ket{m} = \pm \frac{1}{2}$ for all $x$.

\subsection{Infinite chain kink Hamiltonian}

The infinite volume kink Hamiltonian is defined in the standard way via 
the generator of the
Heisenberg dynamics on the algebra of quasi-local observables, 
${\mathcal A}$. This algebra is
the norm closure of the algebra of local observables,
$${\mathcal  A}_{\rm loc} = \bigcup_{\Lambda\subset\mathbb Z, |\Lambda|
  <\infty}{\mathcal  A}_\Lambda, \quad
  {\mathcal  A}_\Lambda = \bigotimes_{x\in\Lambda} {\rm Mat}_{\mathbb C}(2) 
  \,.
$$
We prefer to work in the {\sf GNS} Hilbert space representation of the 
infinite volume kink states.
To this end we introduce the incomplete tensor product Hilbert space
\be\label{defH} {\mathcal  H} = \overline{\bigcup_{\Lambda, |\Lambda|<\infty}
                           \left(\bigotimes_{x\in\Lambda}
                           \mathbb C^{2} \otimes \bigotimes_{y\in\mathbb 
			   Z\mathbb\setminus
			   \Lambda}\Omega(y)\right)} \,,
\ee
where 
$$ \Omega(y) = \left\{\begin{array}{lc} \ket{\uparrow} & \mbox{ if } y\le0\\
                                         \ket{\downarrow} & \mbox{ if } y>0
                 \end{array}\right. \,.
$$		    
Let us define the vector 
\be \Omega = \bigotimes_{y\in\mathbb Z} \Omega(y)\,,
\ee
which is a vector in ${\mathcal  H}$.
We also define the (unnormalized) {\sf GNS} vector
$$ \Omega^{{\sf GNS}} = \sum_{k\ge0}\sum_{x_1<\cdots<x_k\le0<y_1<\cdots<y_k}
                        q^{\sum_{j=1}^k (y_j-x_j)}\prod_{j=1}^k S_{x_j}^1
			S_{y_j}^1 \Omega \,.
$$
The set of vectors $\psi=A\Omega^{{\sf GNS}}$, where $A$ is a local observable, 
constitute a dense 
subspace of $\mathcal  H$. In this representation, the kink Hamiltonian is 
now defined by 
\be\label{def:kink Hamiltonian} 
 H A\Omega^{{\sf GNS}} 
 = \lim_{[a,b]\to\mathbb Z} [H[a,b],A]\Omega^{{\sf GNS}} \,.
\ee

Next we state the assumptions on the magnetic field by which we perturb the 
kink Hamiltonian.

\begin{assumption} \label{assumption}
Let $\Z\times\R\ni(x,t)\mapsto\vec{B}(x,t)\in\R^3$ denote a magnetic field 
which has spatial support in some finite set $\Lambda\subset\Z$, and is 
absolutely continuous with bounded 
derivative (as a function of time). Let $V(t) = \sum_{x\in\Z}\vec{B}(x,t)
\vec{S}_x$. We further impose a scaling on the strength and the variation 
of $V$, and define
\be \label{def:perturbation} V^\lambda (t) = \l V(\l t)\, .
\ee
\end{assumption}

Clearly, $H + V^\l(t)$ is a selfadjoint operator with domain 
${\mathcal  D}(H)$. The reason for our assumption on the support of $B$ 
is that if we want to consider, for simplicity, a time independent and 
translation invariant magnetic field, then, in general, $e^{-it(H+V)}$ is 
not a strongly continuous group of unitaries on ${\mathcal  D}(H)$. The 
application of $V$ on a ground state will no longer be in the Hilbert 
space $\mathcal H$. This is no contradiction to the fact that 
the unperturbed time evolution is strongly continuous in the Heisenberg 
picture, i.e., on the algebra of (quasi-local) observables $\mathcal A$, 
since $V$ would no longer be quasi-local.

We could extend the class in (\ref{assumption}) to perturbations 
relatively bounded to $H$ (with relative bound less than 1 to ensure 
selfadjointness according to the Theorem by Kato-Rellich) but
this does also not allow to consider uniform fields, and we therefore 
stick to bounded perturbations.

We shall first find the leading dynamics, and then consider the limit as 
the spatial support of $\vec{B}(x,t)$ tends to infinity.

\section{Dynamics}\label{dynamics}

\begin{lemma}[Dyson series] Let $U^\l(t,t_0)$ be the solution to the time 
dependent Schr\"odinger equation
 $$ i\frac{d}{dt} U^\l(t,t_0) = (H + V^\l(t)) U^\l(t,t_0)
 $$
with $V^\l$ satisfying (\ref{assumption}).
Then for $\phi\in{\mathcal D}(H)$ and $N\in\mathbb N$ 
\bea \label{eq:Dyson} e^{itH} U^\l(t,t_0) \phi &=& \phi + \sum_{0 < n < N}
(-i)^n \int_{t_0\le t_n\le\ldots\le t_1\le t}d{\bf t} \,
V^\l(t_1,t_1)\cdots V^\l(t_n,t_n)\phi
\nonumber\\
&& + \,\,{\bf E}_N\phi \,,
\eea
where we used the notation $d{\bf t} = dt_1\cdots dt_n$, and $V(t,t) = 
e^{itH} V(t) e^{-itH}$. The error term, ${\bf E}_N$, is bounded by 
$\frac{|t-t_0|^N}{N!} (\l \|V\|)^{N-1}$, which converges
to 0 as $N\to\infty$ uniformly for $t$ on compacts; we use $\|V\|=
\sup\{\|V(s)\|;s\ge t_0\}$.
\end{lemma}

\begin{proof} We may assume that $t_0=0$ and $\l=1$. Setting $\Omega(t) = 
e^{itH} U(t)$, then for $\phi \in {\mathcal D}(H)$ we have $\frac{d}{dt} 
\Omega(t)\phi = -iV(t,t)\Omega(t)\phi$. 
Integrating over $t$ and iterating we obtain
\beax \Omega(t)\phi &=&\phi + \sum_{0<n<N} (-i)^n \int_{0\le t_n\le
\ldots\le t_1\le t} d{\bf t} \, V(t_1,t_1)\cdots V(t_n,t_n)\phi
 \\
&+& (-i)^N \int_{0\le t_{N-1}\le\ldots\le t_1\le t} d{\bf t} \, V(t_{1},t_{1})
     \cdots V(t_{N-1},t_{N-1})\Omega(t_{N-1})\phi\,. 
 \eeax
The error estimate follows then easily.
\end{proof} 

\begin{figure}[h]
\begin{center}
\resizebox{4.5in}{1.25in}{\includegraphics{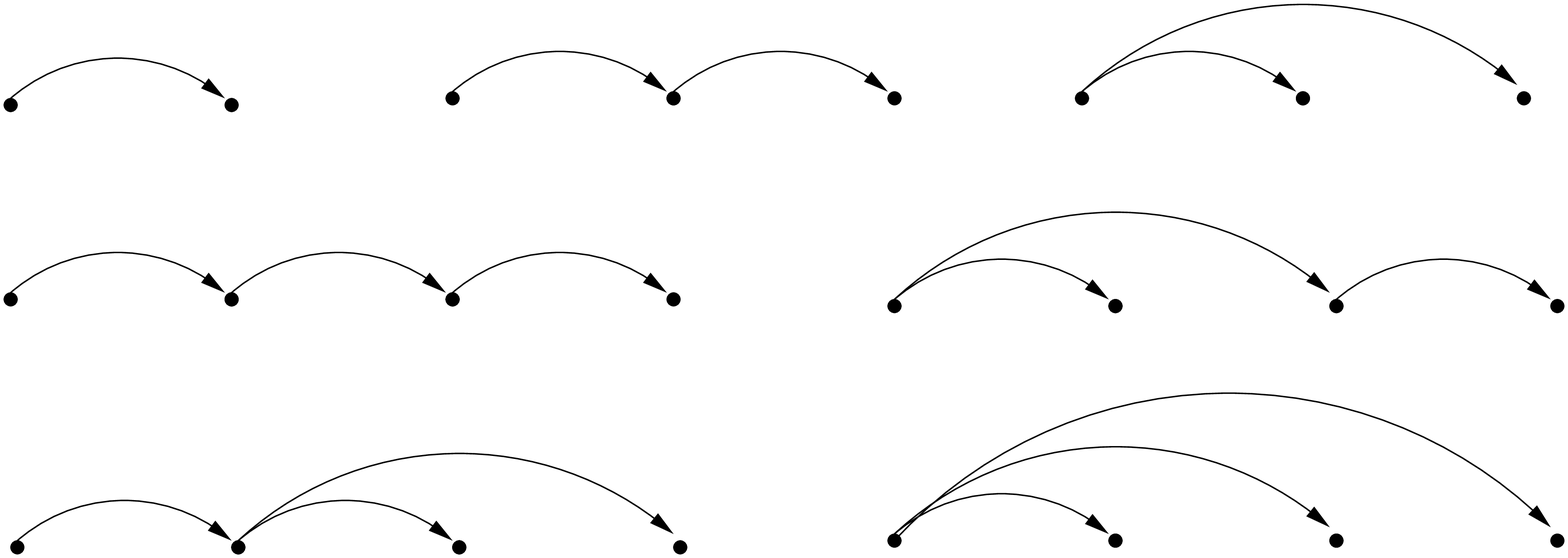}}
\caption{\label{fig:graphs} The first 7 graphs $G$, comprising $\mathcal{G}_1$, 
$\mathcal{G}_2$ and $\mathcal{G}_3$.}
\end{center}
\end{figure}
We discuss next a formula for the iterated $t$ integral allowing us to decide 
which energy terms contribute at what rate. To this end we consider the set 
${\mathcal G}_n$ of directed graphs $G$ with vertex set equal to 
$\{n+1,n,\ldots,1\}$ 
and bonds satisfying the following rules. 1) The direction of a bond is 
from a higher to a lower value of $n$. 2) There is exactly one bond for every 
site $1\le k\le n$, leading to it from a site $p(k)>k$, although many bonds 
can originate from a certain site. We call the bond $b(k)=(p(k),k)$, and 
define $\ell(b(k))=p(k)-k$ the length of this bond,
and $\ell(G)=\sum \ell(b(k))$ the length of the graph $G$. 3) If there is a 
bond between $i$ and $j$, 
then there is no bond possible between $k$ and $l$ if
($k>b(i)$ and $l<b(i)$) or ($k>i$ and $l<i$) or ($k<b(i)$, $l>i$).
In other words, there are no crossing bonds, and no bonds properly nested.
(With these rules, the digraphs $G$ are equivalent to the set of all 
compositions, i.e., all finite ordered sequences of positive integers. 
To a digraph $G=\{b(1),\dots,b(n)\}$,
we associate the composition $[p(1)-1,p(p(1))-p(1),p(p(p(1)))-p(p(1)),
\dots]$. $\mathcal{G}_n$ is the set of all compositions which sum to $n$.)

We can show that $|\mathcal{G}_n|=2^{n-1}$. 
It is clear that $|\mathcal{G}_1|=1$.
If we have a graph $G\in{\mathcal G}_n$, then we may construct two new graphs 
in ${\mathcal G}_{n+1}$. Let us denote the $n+2$nd vertex by 0, then 0 may 
be connected to 1 whose graph we denote by $G_1$.
The only other possibility is to join 0 with $p(1)$ and we call the graph 
$G_2$. We define signs for the graphs, inductively. Let $\curvearrowright$ 
be the unique graph with vertex set $\{2,1\}$
and bond from 2 (left) to 1 (right), then we set $\s(\curvearrowright)=1$. 
Using the above construction, we define the sign by $\s(G_1)=-\s(G)$, and 
$\s(G_2)=\s(G)$. Some graphs are shown in Figure 1.

Given a sequence of numbers $k_j\,:\,1\leq j\leq n$,
for any graph $G\in{\mathcal G}_n$ and vertex $i$ we define $k(i;G) = 
\sum_{i\le j \le i+\ell(b(i))-1}k_j$. 

\begin{lemma}\label{lemma:iterated integrals} 
Let $E_j:1\le j\le n+1$ and $k_j:1\le j\le n$ be two sequences of 
numbers, then
\bea\label{iterated integrals}
\lefteqn{\int_{0\le t_n\le\ldots\le t_1\le t} d{\bf t} \prod_{j=1}^n 
e^{-it_j (E_{j+1}-E_j + i\l k_j)}
=}\nonumber\\
&&\sum_{G\in{\mathcal G}_n}\s({G})\left(\prod_{j=1}^n \frac{i}{E_{p(j)} - 
E_j + i\l k(j;G)}\right) \left(e^{-it(E_{p(1)}-E_1+i\l k(1;G))} -1\right) .
\eea
\end{lemma}

\begin{proof}
We can easily check this by induction. The formula is obviously correct for 
$n=1$. Assuming that 
the formula is true for $n-1$, we may write the $n$ fold integral as 
\beax \lefteqn{\sum_{G\in{\mathcal G}_{n-1}}\s({G})\left(\prod_{j=2}^n
  \frac{i}{E_{p(j)} - E_j + i\l k(j;G)}\right)
}\\
&&\times\int_0^t dt_1 \left(e^{-it_1(E_{p(2)}-E_1+i\l k(2;G)+i\l k_1)} - 
   e^{-it_1(E_2-E_1+i\l k_1)}\right).
\eeax
The vertex set of the graphs $G\in{\mathcal G}_{n-1}$ is here $\{n+1,
\ldots,2\}$. Now, in the first term of the integrand we have that 
$p(2;G)=p(1;H), k(1;H)=k(2;G)+k_1, \s(H)=\s(G)$, where $H$ is 
the graph on $\{n+1,\ldots,1\}$
which agrees on $n+1,\ldots,2$ with $G$ plus a bond connecting 1 with the 
same site as 2. In the second term we have a bond between 1 and 2, and by 
our rules hence a change of the sign.
We can combine this into a sum over all graphs $H\in{\mathcal G}_{n}$, 
which verifies the formula.
\end{proof}

In the following theorem we identify the leading term in the limit when 
$\l\to0$, and $t\to\infty$, such that $\tau=\l t$ is 
constant. Here, $\tau$ can be interpreted as a macroscopic time scale.
We shall use the spectral decomposition of the Hamiltonian
\be
H=\int_0^\infty E\, dP(E)\ ,
\label{res}\ee
such that for any $\phi\in \mathcal H$, we have 
$$
\phi=\int_0^\infty \,dP(E)\phi\ .
$$

\begin{theorem}[Leading dynamics] \label{leading} 
Let $\phi\in\mathcal H$, and let $V$ satisfy 
Assumption (\ref{assumption}). Then, for all $\d$, $1>\d>0$, there
exists a $\l_0$ depending on $\|V\|,\tau,\|\phi\|$ and $\delta$ such that 
for $0<\l<\l_0$
\be \left\|e^{i\lambda^{-1}\tau H}U^\l(\l^{-1}\tau)\phi - \int_0^\infty  
     \mathbb T\left(e^{-i P(E) \int_0^\tau dt\, V(t) P(E)}\right)\, dP(E)\phi 
     \right\|
     < \l^{1-\d} \, .
\ee
$\mathbb T$ means time ordering, i.e., 
$\mathbb T(V(s) V(t)) = V(s) V(t)$ if $s>t$, and zero otherwise.
\end{theorem}

\begin{proof} We may assume that $\|\phi\|=1$.  
Let $N$ be an integer which we choose later. We may write ($t=\l^{-1}\tau$)
\beax
&&\left\|e^{it H}U^\l(\l^{-1}\tau)\phi - \int_0^\infty
     \mathbb T\left(e^{-i P(E) \int_0^\tau dt\, V(t) P(E)}\right)\, dP(E)
     \phi\, \right\|
\\
&&\quad\quad\le|{\bf E}_N| + \left|\Sigma_N-\Sigma_N'\right| + |{\bf E}_N'| \,.
\eeax
Here, ${\bf E}_N$ is the error term from equation (\ref{eq:Dyson}), and 
${\bf E}_N'$ the remainder
term of the power series for the exponential function in 
$\mathbb T\left(e^{-i P(E) \int_0^\tau dt\, V(t) P(E)}\right)\,\phi$ 
to the order $N-1$, whereas $\Sigma_N$ and $\Sigma_N'$ refers to 
their finite sums.

First we take care of the error terms ${\bf E}_N$ and ${\bf E}_N'$ by 
choosing $N=N(\l;\|V\|,\tau)$ such that 
$$ \frac{1}{\|V\|} \frac{(\tau \,\|V\|)^N}{N!}  < \mfr{1}{3}\l^{1-\d}.
$$
This can be accomplished with $N=-C\ln{\l}$ and the constant $C$ depending 
on $\|V\|$ and $\tau$.

Second, we investigate the limit $(\l\to0,t\to\infty,\l t\to\tau)$ for 
the $n$-th term in the finite Dyson series. By inserting the spectral 
resolution for $H$ (\ref{res}), the $n$-th term in 
$e^{itH}U^\l(t)\phi$ equals
\be\label{n-th term} (-i\l)^n \int_{[0,t]^n} d{\bf t}
    \int_{\mathbb R_+^{n+1}}  d{\bf E}_n 
    \prod_{j=1}^n e^{-it_j (E_{j+1}-E_j)} 
\mathbb T\left(\prod_{j=1}^n 
       P(E_j) V(\l t_j)\right) dE_{n+1}\phi \, .
\ee
In order to perform the $t$ integration we write $V$ as its inverse 
Laplace transform. Let $c_1<c_2<\cdots <c_n<0$, then we obtain
\beax \lefteqn{(-i\l)^n \frac{1}{(2\pi i)^n}\int_{c_1-i\infty}^{c_1+i\infty}
\cdots\int_{c_n-i\infty}^{c_n+i\infty} d{\bf k} \int_{\mathbb R_+^{n+1}} 
d{\bf E}_n }
\\
&&
\times\,\int_{[0,t]^n} d{\bf t} \prod_{j=1}^n e^{-it_j (E_{j+1}-E_j + 
i\l k_j)}\,
\mathbb T\left(\prod_{j=1}^n P(E_j) \hat{V}(k_j)\right) dP(E_{n+1})\phi\, .
\eeax
By using formula (\ref{iterated integrals}) we can do the $t$ integration, 
and obtain
\beax \lefteqn{\frac{1}{(2\pi i)^n}\int_{c_1-i\infty}^{c_1+i\infty}\cdots
  \int_{c_n-i\infty}^{c_n+i\infty} d{\bf k} \int_{\mathbb R_+^{n+1}} d{\bf E} 
   \int_{[0,t]^n} d{\bf t} \prod_{j=1}^n e^{-it_j (E_{j+1}-E_j + i\l k_j)}
}
\\
&&
\sum_{G\in{\mathcal G}_n}\s({G})\left(\prod_{j=1}^n \frac{\l}{E_{p(j)} - E_j 
+ i\l k(j;G)}\right)\left(e^{-it(E_{p(1)}-E_1+i\l k(1;G))} -1\right)
\\
&&\times\,\mathbb T\left(\prod_{j=1}^n P(E_j) \hat{V}(k_j)\right) dP(E_{n+1})
\phi\, .
\eeax
Here, it is 
convenient to have $c_i$'s such that there are no zeros in the denominator. 
We observe here the crucial fact that in order to find a term of the order 
1, all $E_j$ have to be equal. We may now go back to
(\ref{n-th term}), change coordinates and conclude that in this limit we get
\beax\lefteqn{ (-i\l)^n \int_0^\infty \int_{[0,t]^n} d{\bf t} \,\mathbb T
 \left(\prod_{j=1}^n P(E){V}(\l t_j)\right) dP(E)\phi
  }\\
  &=&(-i)^n \int_0^\infty \,\mathbb T\left(\int_0^\tau d{t} P(E) V(t)
  \right)^n 
  dP(E)\phi \,. 
\eeax

Finally, we have to bound the sum of all other terms in the finite Dyson 
series which individually tend to 0 as $\l\to 0$,
and are contained in $\left|\Sigma_N-\Sigma_N'\right|$. 
As there are $2^{n-1}$ terms in formula (\ref{iterated integrals}), we 
immediately obtain the bound
$$\left|\Sigma_N-\Sigma_N'\right|\le\frac{\l}{2}\sum_{n=1}^{N-1}
  \frac{(2\tau \|V\|)^n}{n!}\,, 
$$
which in turn is bounded by $\mfr{1}{3}\l^{1-\d}$ if we choose $\l<\l_0$ 
with $\l_0$ such that $e^{2\tau\|V\|} < \frac{2}{3}\l^{-\d}_0$.
\end{proof}

If $H\phi=0$ is a kink ground state of $H$, then the leading term equals
\be\mathbb T\left(e^{-i \int_0^\tau P(0) V(t) P(0)}\right)\phi \,.
\ee 
This expression will be further analyzed for a specific choice of $V$ in 
Section~\ref{leading dynamics}. 

Note that so far we did not use any specific 
spectral properties of the Hamiltonian, and thus Theorem \ref{leading} is 
valid, for instance, for any value of the spin. 

We know that $\sigma(H) \subset \{0\} \cup [1-\Delta^{-1},\infty)$, 
using the theorem of \cite{KN1} that there is a gap of $1-\Delta^{-1}$ 
above the infinite volume ground states. If we decompose the kink 
Hamiltonian on $\Z$ as the sum of a kink plus a droplet 
Hamiltonian, i.e., $H^{+-}_\Z = H^{+-}_{(-\infty,y]} + H^{--}_{(y,\infty)}$ 
with $y\in\Z$, then we get approximate 
eigenvectors by tensoring a kink ground on $(-\infty,y]$ (and with interface 
far to the left of $y$) with eigenvectors of $H^{--}_{(y,\infty)}$. In the 
case $y=-\infty$ the spectrum of the latter has
been shown by Babbitt and Gutkin~\cite{BG} to be $\{0\}\cup [1-\D^{-1},
\infty)$. 
Their analysis follows the earlier work by Babbitt 
and Thomas~\cite{BT} for $\D=1$ proving the completeness of the Bethe 
ansatz eigenfunctions in the infinite volume limit and thereby also 
showing that the spectrum is absolutely continuous. 

This way, one obtains $\sigma(H) = \{0\} \cup [1-\Delta^{-1},\infty)$.
But this construction does not give the nature of the spectrum in the interval
in $[1-\Delta^{-1},\infty)$. We plan to address this issue in the future; 
for now we state the following hypothesis. 

\begin{hypo} \label{conject} 
The spectrum of the spin-$\frac{1}{2}$ kink Hamiltonian $H$ with $\D>1$ has 
no other point spectrum than 0, and the rest is absolutely continuous,
equal to $[1-\D^{-1},\infty)$. 
\end{hypo} 

By assuming this information we can identify the first correction to the
leading term given in Theorem \ref{leading} for ground states, and we show that
it is ${\mathcal O}(\l)$. For simplicity, we state the result only for time
independent fields and leave the general case to the reader. The main
argument in the proof below, however, is carried out for a time dependent 
field. 

\begin{theorem}[First order correction] \label{thm:correction}
Let $\phi$ be a (normalized) ground state of the spin-$\frac{1}{2}$ kink 
Hamiltonian, $V$ a perturbation as in Assumption (\ref{assumption}) and 
$H$ satisfying Hypothesis (\ref{conject}). 
Then,
\bea \label{eq:correction}
\lefteqn{\left\|e^{i\l^{-1}\tau H} e^{-i\l^{-1}\tau(H+\l V)}\phi - 
e^{-i\tau P(0)V}\phi - \l \int_{1-\D^{-1}}^\infty \frac{dE}{E} P(E) 
V e^{-i\tau P(E) V}\,\phi\right.
}\nonumber\\
&&-\left. P(0)V \left(e^{-i\tau [P(0) V + \l H^{-1} ({\bf 1} - P(0))V]} - 
e^{-i\tau P(0)V}\right)\phi\right\|  = o(\l)\, .
\eea
\end{theorem}
\begin{proof} We analyze the $n$-th term in the finite 
Dyson series and treat the error terms as before by choosing first 
$N=N(\l,\|V\|,\tau)$.
The spectral resolution of $H$ has now the form ${\bf 1}= P(0) + 
\int_{1-\D^{-1}}^\infty dP(E)$. We have seen above that to leading order 
all energies $E_i$ have to be equal to 0. According to 
(\ref{iterated integrals}), terms of the order $\l$
come from sequences $(E_m-E_l):\{m<l\}\in{G}$ such that all but one pair 
is non-zero. We represent such sequences graphically by polygons with 
$n+1$ vertices. We draw a horizontal line between the vertices $k$ and 
$k+1$ whenever $E_k = E_{k+1}$, an upward
line if $E_k>E_{k+1}=0$, and a downward line if $0=E_k<E_{k+1}$. 

By our rules for the graphs $G\in{\mathcal G}_n$, the only possible 
polygons contributing to the order ${\mathcal O}(\l)$ are of the form 
displayed in Figure~\ref{fig:polygon2}.
\begin{figure}[h]
\resizebox{4.5in}{1.25in}{\includegraphics{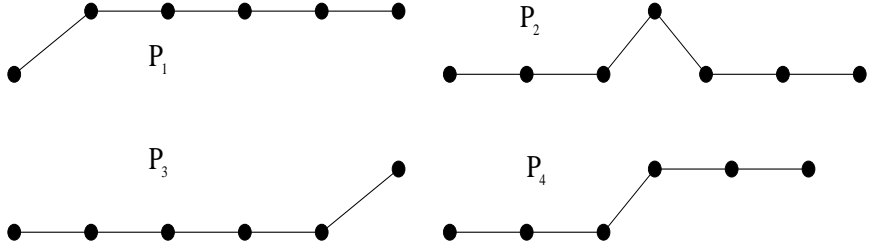}}
\caption{\label{fig:polygon2} Sequences of energies satisfying the 
condition that their corresponding differences is non-zero for exactly 
one pair. The bump in polygon $P_2$ can be at any site such that the 
starting and ending point of the polygon lies at the horizontal line. 
}
\end{figure}

Using Hypothesis \ref{conject} we shall show that, in fact, only $P_1$ 
and $P_2$ contribute to ${\mathcal O}(\l)$ while $P_3$ and $P_4$ are of 
lower order. We shall now consider the contributions
from $P_1,P_2,P_3$ and $P_4$ to formula
(\ref{n-th term}). Let us start with polygon $P_1$, i.e., we have a sequence 
$0=E_{n+1}<E_n =\ldots=E_1=E$. 
In
\beax\lefteqn{\int_0^{t_{n-1}} dt_n\, e^{it_n E} V(\l t_n)\phi
=}\\
&&\frac{1}{iE} \left( e^{it_{n-1}E} V(\l t_{n-1}) - V(0)\right)\phi -
   \frac{1}{iE} \int_0^{t_{n-1}} dt_n\, e^{it_n E} V(\l t_n)\phi
\eeax
we can use that, by our assumption on the spectrum of $H$ and the 
Riemann-Lebesgue Lemma, the leading term is $\frac{i}{E}V(0)$, and the 
rest is $o(1)$, where $o(1)$ depends on the inverse gap
$\D/(\D-1)$ and $\sup\{\|V(s)\|\,:\, s\in[0,\tau]\}$. Then, up to $o(1)$ 
we have
\beax \lefteqn{\l \int_{1-\D^{-1}}^\infty dE\, \frac{1}{E}
(-i\l)^{n-1} \int_{0\le t_{1}\le\cdots\le t_{n-1}} d{\bf t} 
\prod_{j=1}^{n-1} P(E) V(\l t_j) P(E)V(0)\phi
}\\
&=&\l\int_{1-\D^{-1}}^\infty dE\, \frac{1}{E} (-i)^{n-1} \mathbb T 
\left(\int_{0}^\tau
d\tau_1 P(E) V(\tau_1)\right)^{n-1} P(E) V(0)\phi \,,
\eeax
where we have used a change of variables as in the proof of Theorem 
\ref{leading}. This gives the
first correction term in (\ref{eq:correction}).

Next, we consider the polygon $P_2$ with all energies equal to 0 except
for $E_m=E>0$ with $2\le m\le n$. Then, the contribution of $P_2$ to the 
$n$-th term in the Dyson series is equal to
\beax (-i\l)^{m+1} \int_{1-\D^{-1}}^\infty dE\,\int_{0\le t_m\le\cdots\le 
t_1\le t} d{\bf t}_m e^{-it_{m-1} E +it_m E} \left[\prod_{j=1}^{m-1} P(0) 
V(\l t_j)\right]\, A(\l t_m)
\eeax      
with 
$$A(\l t_m) = P(E)V(\l t_m) (-i\l)^{n-m-1} \int_{0\le t_n\le\cdots\le t_m} 
d{\bf t}_{n-m}\prod_{j=m+1}^{n} P(0) V(\l t_j)\phi\,.
$$
Note that $A(0)=0$. In order to simplify matters, we let $V$ be time 
independent and perform the
time integration. Hence, we get (to leading order)
\beax \lefteqn{-\l \frac{(-i\l t)^{n-1}}{(n-1)!}\int_{1-\D^{-1}}^\infty dE
\,\frac{1}{E}\Big(P(0)V\Big)^{m-1}\,P(E) V\,\Big(P(0) V\Big)^{n-m}\phi
}\\
&=&\l\frac{(-i\tau)^n}{(n-1)!} \Big(P(0) V\Big)^{m-1} 
\,H^{-1} ({\bf 1}- P(0)) V \, \Big(P(0) V\Big)^{n-m}\phi \, .
\eeax
Now we can easily sum up the contributions of all $n-2$ polygons of the 
form $P_2$, and obtain the $n$-th term in the power series of the 
exponential in the second correction term in
(\ref{eq:correction}) up to errors $o(\l)$. 

Finally, we prove that the other two kinds of polygons contribute of lower 
order. For the polygon $P_3$, (\ref{n-th term}) equals
$$(-i\l) \int_0^t dt_1\,e^{it_1 E} P(E) B(\l t_1)\,,
$$
with 
$$B(\l t_1) = (-i\l)^{n-1} \int_{0\le t_n\le\cdots\le t_1} d{\bf t}\,V(\l t_1)
  \prod_{j=2}^n P(0)V(\l t_j)\phi\,.
$$
$B(0)=0$, and by integration by parts we get
\beax (-i\l) \int_0^t dt_1\,e^{it_1 E} P(E) B(\l t_1)
=-\frac{\l}{E} e^{it E} P(E) B(\l t) + o(\l)
\eeax
by our assumption on $H$. 

For a sequence of the form $P_4$, with $E=E_1=\ldots=E_m>E_{m+1}=\ldots=
E_{n+1}=0$. Then, the expression in formula (\ref{n-th term}) equals
$$(-i\l)^m \int_{0\le t_m\le \cdots\le t} d{\bf t} \, e^{it_mE} \prod_{j=1}^m 
P(E) V(\l t_j)C(\l t_m)\, ,
$$
with 
$$C(\l t_m) = (-i\l)^{n-m} \int_{0\le t_n\le \cdots\le t_m} d {\bf t} 
\prod_{j=m+1}^n P(0) V(\l t_j) \phi\,.
$$
Integration by parts in the first integral shows that this term is $o(\l)$; 
in fact, if we assumed $V$ to be $m$ times differentiable, then it would be 
$o(\l^m)$.
\end{proof}

Notice also that the first correction term of the order $\l$ contains the 
information on the whole
spectrum while the leading term only depends on the highly reduced space of
ground states, which we analyze in the next Section.

\section{Leading dynamics}\label{leading dynamics}

In this Section we consider the leading dynamics generated by $P(0)VP(0)$. 
We start with a uniform magnetic field $V_\L$ for which we  
determine an explicit expression as $\L\to\infty$. 

\subsection{Uniform time-independent magnetic field}

Let $\vec{B}(x,t)=\vec{B}$ be a constant vector, and thus $V_\L = 
\sum_{x\in \L}\vec{B}\cdot\vec{S}_x$. We shall consider $K_\L = P(0) 
V_\L P(0)$ as a Jacobi operator (i.e., tridiagonal) on $\ell^2(\Z)$. 
Since $H$ from (\ref{H}) commutes with global rotations in the $xy$ 
plane, we choose a new basis $|n,U\rangle = U^{-1}|n\rangle$ of $P(0)
\mathcal H$, where $U$ satisfies $UV_\L U^{-1} = \sum_{x\in\L}
\sqrt{B_1^2+B_2^2} S_x^1 + B_3 S_x^3$. 
In other words, $U=\prod_x e^{i \theta S_x^3}$, where 
$(B_1,B_2) = (\sqrt{B_1^2+B_2^2} \cos\theta,\sqrt{B_1^2+B_2^2} \sin\theta)$.

When taking the infinite volume limit we have to renormalize the total 
$S^3$ component, and define
\be S^3 = \lim_{n\to\infty} \sum_{|x|\le n} S^3_x \,.
\ee 
In the new basis, we have
\beax\lefteqn{\langle m,U|K_\L|n,U\rangle}
\\
&=& \langle m|P(0) U V_\L U^{-1} P(0)|n\rangle
\\
&=&\sum_{x\in\L}\sqrt{B_1^2+B_2^2} \langle m|[S_x^+ + S_x^-]|n\rangle + B_3
\langle m|S_x^3|n\rangle
\\
&=&\sqrt{B_1^2+B_2^2}\, a_\L(n)\, (\d(m,n-1) + \d(m,n+1)) + B_3 b_\L(n) 
\d(n,m)\,,
\eeax
where the index $n\in\Z$ denotes the eigenvalue of the renormalized total 
$S^3$ compo\-nent. In Appendix A we calculate $a=\lim_{\L\to\Z}a_\L(n)$, 
see equation (\ref{a1}). Notice that the limit is independent of $n$ by 
translation invariance. 
The function $b_\L(n)$ satisfies $|b_\L(n)| \leq |\L|$, and
$\lim_{\L \to \Z} b_\L(n) = n$ for all $n$.

For finite $\L$, $K_\L$ is a trace class operator, and the spectrum is 
pure point, accumulating to zero.
We now discuss the case when $\L=\Z$.

Let $\D$ be the lattice Laplacian with unit off-diagonal entries and 
zero diagonal terms on $\ell^2(\Z)$, and let $W(n)=n$ be the Stark potential. 
Then we define the Stark-Jacobi operator,
\be\label{Stark-Jacobi} K_0 = \a \D + \c W
\ee
with $\a=\sqrt{B_1^2+B_2^2}a$ and $\c=B_3$. 

The operator $K_0$ is selfadjoint with dense domain ${\mathcal D}(K_0) = 
\{f\in \ell^2(\mathbb Z) : (nf(n))_n \in \ell^2(\mathbb Z)\}$. This follows 
from the fact that $W$ is selfadjoint on ${\mathcal D}(K_0)$, and that 
$\D$ is bounded and selfadjoint. 

As for the continuous case, one can find the spectrum explicitly but it turns
out to have rather different properties.
Only in the case $\c=0$ we shall have pure absolutely continuous spectrum 
which is equal to $[-2\a,2\a]$. As soon as $\c\not=0$, the spectrum is pure
point. The latter is analyzed in the following lemma, and the generalization 
to higher dimensions is given in Appendix B.

\begin{lemma} Let $K_0 = \a\D +\c W$ be the Stark-Jacobi operator 
(\ref{Stark-Jacobi}), and $\c\not= 0$. Then, $\textrm{spec}(K_0)=\c 
\mathbb Z$ is pure point spectrum with simple eigenvectors 
$\phi_m(n) = J_{m-n}(2\a/\c)$; $J_n$ denotes the (ordinary) Bessel function 
of order $n$. 
\end{lemma}

\begin{proof} 
By Fourier transforming the eigenvalue equation
$\a\phi(n-1) + \a\phi(n+1) + \c n \phi(n) = \lambda \phi(n)$, we obtain
$$ \left(2\a \cos k + i\c \frac{d}{dk}\right) \phi(k) = \lambda \phi(k),
$$
where $k\in[-\pi,\pi]$. This is solved by $\phi(k) = e^{i(2\a\sin{k} -
\lambda k)/\c}$. The periodicity condition $\phi(k+2\pi) = \phi(k)$ shows 
that $\lambda\in\c\mathbb Z$. Note, that
$\sum_{m\in\mathbb Z} \phi_m(k) = 2\pi \d(k)$ which proves that these 
(mutually orthogonal) eigenfunctions are complete. In configuration space 
we thus have
\beax \phi_m(n)& =& \frac{1}{2\pi} \int_{-\pi}^\pi 
                    \cos{\left(\mfr{2\a}{\c} \sin{k} + k(n-m)\right)}\, dk
      = J_{n-m}\left(\mfr{2\a}{\c}\right) \,.
\eeax
\end{proof}

\begin{lemma} The kernel for the reduced dynamics generated by $P(0) V P(0)$, 
and hence of $K_0$, has the following form: 
\bea \langle x,U|e^{-itP(0)V_\Z P(0)}|n,U\rangle &=& \Big[e^{-iK_0 t}\d_n
\Big](x) \nonumber\\
&=&J_{n-x}\left(\mfr{4\a}{\c}\sin{(\mfr{\c t}{2})}\right)
    \exp\left[-i \Big( \frac{\c t - \pi}{2} x + \frac{\c t + \pi}{2} n 
    \Big)\right]\,.
\eea
\end{lemma}

\begin{proof} 
\beax e^{-iK_0 t}\d_n(x)&=&\sum_{m\in \mathbb Z} e^{-i\c m t} \phi_m(x) 
\langle\phi_m|\d_n\rangle
\\
&=&\sum_{m\in \mathbb Z} e^{-i\c m t} J_{x-m} (\mfr{2\a}{\c}) J_{n-m} 
(\mfr{2\a}{\c})
\\
&=&\sum_{m\in \mathbb Z} e^{-i\c n t} e^{i\c m t} J_{m-n+x} (\mfr{2\a}{\c}) 
J_{m} (\mfr{2\a}{\c})
\\
&=& e^{-i\c n t} e^{i (\pi - \c t) (x-n)/2} J_{n-x}\left(\mfr{4\a}{\c} 
\sin{(\mfr{\c t}{2})}\right).
\eeax
The last step is content of Graf's addition theorem, see 
\cite{AS}, formula (9.1.79). 
\end{proof}

For future ease of notation, we define
$w = \mfr{4\a}{\c}\sin{(\frac{\c t}{2})}$, and 
$\chi = \frac{\pi-\c t}{2}$. 

We now inspect closer the reduced time evolution of the magnetization 
profile of a kink ground state localized at 0. Defining 
$$
m(\hat{\Omega},x,t) = \langle 0,U|e^{itP(0)VP(0)}(\hat{\Omega} \cdot 
\vec{S}_x)e^{-itP(0)VP(0)}|0,U\rangle
$$
for any vector $\hat{\Omega} \in \mathbb{S}^2$, we calculate
\beax \lefteqn{m(\hat{\Omega},x,t)}\\
&=&\sum_{m,m'\in \Z}
   \langle 0,U|e^{itP(0)VP(0)}|m,U \rangle\,\langle m,U|\hat{\Omega} 
   \cdot \vec{S}_x^3|m',U\rangle \, 
   \langle m',U|e^{-itP(0)VP(0)} |0,U\rangle 
\\
&=&\sum_{m,m'\in \Z} e^{i \chi (m'-m)} J_m(w) J_{m'}(w) 
   \langle{m | R_{-\theta}(\hat{\Omega}) \cdot \vec{S}_x | m'}\rangle\, ,
\eeax
where $R_{-\theta}$
is rotation about the $z$ axis by an angle $-\theta$.
(Note this is the opposite rotation as that used to go from $P(0) V P(0)$ 
to $K_0$.) Taking $\hat{\Omega} = \hat{e}_3$,
and defining $m^3(x,t) = m(\hat{e}_3,x,t)$, we have
\beax m^3(x,t)&=&
\sum_{m,m'\in \Z}  e^{i \chi (m'-m)} J_m(w) J_{m'}(w) 
   \langle{m | S_x^3 | m'}\rangle \\
&=&\sum_{m\in \Z} J_m^2(\mfr{4\a}{\c}\sin{(\mfr{\c t}{2})})\, 
   \langle m|S_x^3|m\rangle .
\eeax
Similarly,
defining $m^1(x,t) = m(\hat{e}_1,x,t)$, and observing that
$R_{-\theta}(\hat{e}_1) = a \hat{e}_1 - b \hat{e}_2$,
where $a = \cos \theta = B_1/\sqrt{B_1^2 + B_2^2}$
and $b = \sin \theta = B_2/\sqrt{B_1^2+B_2^2}$, we have
\beax m^1(x,t)
&=& \sum_{m,m'\in \Z} e^{i \chi (m'-m)} J_m(w) J_{m'}(w) 
   \langle m|[a S_x^1 - b S_x^2]|m' \rangle  \\
&=& \sum_{m,m'\in \Z} e^{i \chi (m'-m)} J_m(w) J_{m'}(w) 
   \langle{m | \left[ \frac{a+ib}{2} S_x^+ + \frac{a-i b}{2} S_x^- \right] 
   | m'}\rangle \,.
\eeax
We now observe a phase condition on $|m\rangle$ that guarantees
$\langle{m|S^-|m'}\rangle$ is real. This means that
\beax \lefteqn{m^1(x,t)}\\
  &=& \sum_{m,m'\in \Z}  e^{i \chi (m'-m)} J_m(w) J_{m'}(w) 
   \left(\frac{a-ib}{2} \langle{m|S_x^-|m'}\rangle 
  + \frac{a+ib}{2} \langle{m'|S_x^-|m}\rangle\right) \\
&=& \left[\frac{e^{i\chi}(a-ib)}{2} + \frac{e^{-i\chi}(a+ib)}{2}\right]
  \sum_{m\in \Z} J_m(w) J_{m+1}(w) \langle{m|S_x^-|m+1}\rangle\\
&=& [a \cos(\chi) + b \sin(\chi)]
  \sum_{m\in \Z} J_m(w) J_{m+1}(w) \langle{m|S_x^-|m+1}\rangle\\
&=& [a \sin(\mfr{\gamma t}{2}) + b \cos(\mfr{\gamma t}{2})]
  \sum_{m\in \Z} J_m(\mfr{4\a}{\c}\sin{(\mfr{\c t}{2})}) 
  J_{m+1}(\mfr{4\a}{\c}\sin{(\mfr{\c t}{2})}) \langle{m|S_x^-|m+1}\rangle\, .
\eeax

Notice that for $\g\not=0$ the motion is periodic whose period $2\pi \g^{-1}$
is solely 
determined by the field in the $z$ direction. The profile $m^3(x,t)$ is given
by an absolutely convergent series. Therefore, pointwise convergence implies
convergence on compacts. As the  function is periodic in $t$ and thus the 
image is a compact interval, the profile stays  exponentially localized for 
all times.

However, as $\g\to0$ 
\bea
\label{m3eq}
 m_{\g=0}^3(x,t)&=&\sum_{m\in \Z} J_m^2(2\a t)\, \langle m|S_x^3|m\rangle ,
\\
\label{m1eq}
m_{\g=0}^1(x,t)&=& b\sum_{m\in \Z} J_{m}(2\a t)J_{m+1}(2\a t)
\, \langle m|S_x^+|m-1\rangle \,.
\label{m_1}
\eea
Next, we analyze $m_{\g=0}^3(x,t)$ in more detail. A very similar situation 
has been studied by Antal et al~\cite{ARRS} in the context of the time 
evolution of the {XX} model. We follow their analysis. Although we have 
just derived explicit formulas for the profiles, which can be 
analyzed numerically, we prefer to have simpler expressions. As we shall 
show, for large $x$ and $t$, $m^3(x,t)$ will be a function of the velocity 
$v=\frac{x}{t}$, only. To this end  
we define the discrete derivative of $m^3(x,t)$ as
\beax\phi'_x(v) &:=&t\Big[m_{\g=0}^3(x+1,t)-m_{\g=0}^3(x,t)
\Big]_{\frac{x}{t}=v}
\\
&=&\mfr{x}{v} \sum_{m\in\Z} J^2_m(\mfr{x\a}{v}) \,
   \langle 0|S_{x-m-1}^3 - S_{x-m}^3|0\rangle \,,
\eeax   
and study the limit $\lim_{x\to\infty}\phi'_x(v)$. 
Note that $\langle 0|S_{x-m-1}^3 - S_{x-m}^3|0\rangle \geq 0$ for all $x$ 
and $m$ because the profile of the kinks is a decreasing function.
If we define $p(m) = \langle 0|S_{m-1}^3 - S_{m}^3|0\rangle$,
then since it is a telescoping sum
$$
\sum_{m=-\infty}^\infty p(m) = \lim_{m\to \infty}\langle 0|S_{m}^3|0\rangle 
  - \lim_{m\to -\infty}\langle 0|S_{m}^3|0\rangle = 1\, ,
$$
i.e., $p(m)$ is a probability distribution.
All moments of $p$ are finite and the first moment of $p$ is $\mfr{1}{2}$.
In fact $p(m) \sim e^{-c|m|}$ for some positive $c$ depending on $\Delta>1$
(see Lemma A.1.1).
Thus,
$$
\phi_x'(v) = -\sum_{m \in \Z} \mfr{x}{v} J_{m+x}^2(\mfr{x\alpha}{v}) p(m)\, .
$$
We used the fact that $J_{-n}(x)^2 = J_n(-x)^2 = J_n(x)^2$.
One may easily bound $\mfr{x}{v} J_{x-m}^2(\mfr{x\alpha}{v})$ uniformly in $x$
and $m$, which our asymptotics will show (see below). 
Hence, we may interchange the limits $x\to\infty$ and the sum, 
to calculate 
$$
\lim_{x\to\infty} \phi'_x(v) = \sum_{m\in\Z} p(m) 
  \lim_{x\to\infty} \mfr{x}{v} J_{m+x}^2(\mfr{x\alpha}{v})\, .
$$

We distinguish between the two cases $v > 2\a$, and $v < 2\a$. 
The asymptotic behavior will be very different in these two cases. Let us 
start with $v>2\a$. Then by (9.3.2) in \cite{AS}, asymptotically for large 
$x$
\beax2\mfr{x\a}{v} J^2_{m+x}(\mfr{2x\a}{v}) &=& \mfr{x}{m+x}\, 
\mfr{2(m+x)\a}{x} J^2_{m+x} \left((m+x)\mfr{2x\a}{(m+x)v}\right)
\\
&\sim&\frac{1}{2\pi\sqrt{\mfr{v^2}{4\a^2}-1}} e^{-\frac{2x\a}{v}(1 - 
\sqrt{\frac{v^2}{4\a^2}-1})} \,.
\eeax
Since $p(m)$ is a probability measure, which decays exponentially,
the bound above shows that $\phi'_x(v)$ is exponentially small, and 
$\lim_{x\to\infty}\phi'_x(v) = 0$, which implies that $\phi(v):=
\lim_{x\to\infty}\phi_x(v)=-\frac{1}{2}$. This behavior is not 
surprising for it 
just means a finite speed of propagation, which has to be less than $2\a$. 

However for $0\leq v<2\a$, the behavior of the Bessel functions changes, 
cf (9.3.3), \cite{AS}:
$$\mfr{2x\a}{v} J^2_{m+x}(\mfr{2x\a}{v}) \sim \frac{2}{\pi
\sqrt{1-\mfr{v^2}{4\a^2}}}\cos^2{\Big(x\sqrt{\mfr{\a^2}{4v^2} - 1} - 
x\arccos{(\mfr{v}{2\a})} -\mfr{\pi}{4}\Big)} \,.
$$
This implies that for large $x$
\beax 
\phi'_x(v) &\sim&\frac{1}{\a\pi\sqrt{1-\mfr{v^2}{4\a^2}}}  
\cos^2{\Big(x\sqrt{\mfr{4\a^2}{v^2} - 1} - x\arccos{(\mfr{v}{2\a})} -
\mfr{\pi}{4}\Big)}\, .
\eeax
For $v<2\a$ we recover the function $\phi(v)$ by integration, 
(we switch the limit and integral by dominated convergence,
since $\phi'_x(v)$ is uniformly bounded in $x$ and $v$ for $v\in 
[0,\mfr{2}{\a}-\epsilon]$), i.e.,
\beax \phi(v)&:=&\lim_{x\to\infty}\phi_x(v)
=\int_0^{v} dy\, \lim_{x \to \infty} \phi'_x(y)
=-\frac{2}{\pi} \arcsin{(\mfr{v}{2\a})} \,.
\eeax
The oscillating cosine contributes an average factor $\frac{1}{2}$ to the 
integral. Using symmetry, and collecting terms, we have shown that 
\be\lim_{t\to\infty : x=v t} m_{\g=0}^3(x,t) =\phi(v) = \left\{
\begin{array}{ccc} +\mfr{1}{2} &\mbox{for}&v<-2\a\\
   -\frac{2}{\pi} \arcsin{(\mfr{v}{2\a})}&\mbox{for}&-2\a<v<2\a\\
   -\mfr{1}{2} &\mbox{for}&v>2\a \end{array}
\right.\,.
\ee

The importance behind a further analysis of the profile perpendicular to 
the $z$ direction is that it allows us to decide whether the state is 
rotated by the perturbation into the $xy$ plane or 
whether there is ballistic diffusion of the interface. 
Comparing equations (\ref{m3eq}) and (\ref{m1eq}),
we see that the main difference is a replacement of the probability measure
$p(m) = \langle 0|S_{m-1}^3 - S_{m}^3|0\rangle$ by the signed measure
$\tilde{p}(m)= \langle 1|S_{m}^+ - S_{m-1}^+|0\rangle$.
One can easily determine that $|\tilde{p}(m)| < C_1 e^{-c|m|}$ 
(see Lemma A.1.2). However, now $\tilde{p}(m)$ oscillates, and again 
because of a telescoping sum,
$$
\sum_{m=-\infty}^\infty \tilde{p}(m) = \lim_{m\to \infty} \langle 1|
S_{m}^+|0\rangle - \lim_{m \to -\infty} \langle 1|S_{m}^+|0\rangle
  = 0-0=0\, .
$$
Since we have determined that for any fixed finite $m$,
$\lim_{x \to \infty} \mfr{x}{v} J^2_{m+x}(\mfr{2x\a}{v})$ is a fixed 
function of $v/2\a$ independent of $m$, and in 
particular this implies the same limit for
$\lim_{x \to \infty}\mfr{x}{v} J_{m+x}(\mfr{2x\a}{v}) J_{m+x+1}
(\mfr{2x\a}{v})$, the oscillation of $\tilde{p}$ implies that 
\beax
\lim_{x \to \infty, x=vt} \psi'_x(v) &:=&t\Big[m_{\g=0}^1(x+1,t)-
     m_{\g=0}^1(x,t)\Big]_{\frac{x}{t}=v}
\\
&=&\lim_{x \to \infty} 
b \sum_{m\in\Z} \tilde{p}(m)\lim_{x \to \infty} 
\mfr{x}{v} J_{m+x}(\mfr{2x\a}{v}) J_{m+x+1}(\mfr{2x\a}{v})\\
&=& 0\,.
\eeax   
This does not say that the profile in the $xy$ plane is vanishing. 
Quite on the contrary, it means that it stays exponentially 
concentrated around the kink at 0, and does not follow the ballistic
motion of the third component.

Let us shortly comment on time dependent uniform fields with time 
dependent Stark-Jacobi operator
$K(t) = \a(t)\D + \g(t) W$. It is easy to see that if $B_3(t)=0$
then we just have to replace $\a t$ by $\int_0^t \a(s)\, ds$. E.g.,
$$ m^3(x,t) = \sum_{m\in \Z} J_m^2(2\mbox{$\int_0^t$} \a(s)\, ds)\, 
\langle m|S_x^3|m\rangle \,.
$$ 
We have not been able to derive explicit expressions in the general case, 
but one can show, for instance, by repeated use of Graf's addition theorem, 
that there is a function $F$ such that 
$$ m^3(x,t) = \sum_{m\in \Z} J_m^2(F(\a(t),\g(t))\, \langle m|S_x^3|m
  \rangle \,.
$$ 
In these time dependent cases, a scaling limit as before does not exist.

\subsection{General infinitely extended fields}

Since a thorough discussion would be out of the scope of this paper, we only
want to convince the reader that the leading dynamics of the ``many-body''
Hamiltonian leads to a well studied one-body problem which has seen great
progress in recent years.

Let us start with the simplest case when the magnetic field $\vec{B}(n)$ is 
asymptotically uniform with non-zero third component. I.e., we assume that 
there is a vector $\vec{B}$ and some $p>1$ such that 
$(B_3(n)-B_3)n\in\ell^p(\Z)$, and $(B_{1,2}(n)-B_{1,2})\in\ell^p(\Z)$. Let 
$K$ denote the Jacobi operator corresponding to this vector field. Then $K$ 
is a compact perturbation of the Stark-Jacobi operator $K_0$ in 
(\ref{Stark-Jacobi}). By Weyl's Theorem, $K$ has pure point spectrum
and the motion is quasi-periodic and the interface stays localized at 0.

The case of a perturbation of a uniform field with $B_3=0$ is much more 
complicated. If we perturb by a field in the $z$ direction, which 
corresponds to adding a potential to the (discrete) 
Laplace operator, then we are in the widely studied case of a (discrete) 
Schr\"odinger operator. 

The effect of a perturbation of $\vec{B}$ in the $xy$ plane, however, is
non-local. I.e., if we change $B_1$ at 0 into $B_1(0)$, then the off-diagonal 
matrix element computed from 
$$\bra{n}\sum_{x\in\Z} B_1(x) S_x^+\ket{n-1} = B_1\bra{n}\sum_{x\in\Z}  
  S_x^+\ket{n-1} +(B_1(0) -B_1) \bra{n} S_0^+\ket{n-1}
$$
is effected for all $n$, although exponentially localized at the impurity 
at 0. By a rotation, one can map this Jacobi operator into a Schr\"odinger 
operator. Any non-uniform perturbation of the uniform case by a bounded 
perturbation will modify the spectrum of the reduced Jacobi 
operator, as was recently proven by Killip and Simon~\cite{KS}.

There are many more interesting fields for which we would like to understand 
the time evolution of interfaces. A particularly simple case is a sharply 
localized field at a single site $y$, i.e., $V=\vec{B}\cdot\vec{S}_y$. 
Here, one should rather approximate the time evolution by the projection of 
$H+\vec{B}\cdot \vec{S}_y$ onto its low energy spectrum which seems to be 
separated by a gap from the rest of the spectrum, and 
which is similar to the spectrum of the unperturbed kink Hamiltonian. 
So far only the ground state and the gap above the ground state of this 
Hamiltonian have been studied by Contucci, Nachtergaele and 
Spitzer~\cite{CNS}. We plan to pursue this in the future.

\begin{appendix}

\section{Magnetization profiles}

The exact formula for the spin-$\frac{1}{2}$ magnetization profile of kink 
states in the $z$ direction
was derived in~\cite{St2}, pp 55--58. Note that by rotational symmetry, 
$\langle n|S^1_x|n\rangle =\langle n|S^2_x|n\rangle =0$. We introduce the 
function
\be \label{function f}
    f(z) = \sum_{k\ge0} (-1)^k z^k q^{k(k-1)} .
\ee

\begin{lemma} For the spin-$\frac{1}{2}$ infinite chain kink Hamiltonian 
we have the following profiles of the normalized kink state, $\ket{0}$ 
(which has total $z$ component 0):
\begin{enumerate}
\item If $x>0$, then
\be \langle 0|S^3_x|0\rangle = -\frac{1}{2} + q^{2x}\sum_{k=0}^\infty
(-1)^k q^{k(k+2x+1)}\,,
\ee
while for $x\le0$ we have $\langle 0|S^3_x|0\rangle = -\langle 0|
S^3_{1-x}|0\rangle$. Further,
$p(m)=\bra{0}S^3_{m-1}-S^3_m\ket{0}\le e^{-c|m|}$ with $c$ depending on $q$.
\item $\langle n|S^-_0|n-1\rangle = q^{|n|} f(q^{2|n|+2})$, and $\tilde{p}(m)=
\langle 1|S_{m}^+ - S_{m-1}^+|0\rangle$ satisfies $|\tilde{p}(m)| < 
C_1 e^{-c|m|}$ with $c$ depending on $q$. 
\item The number $a$ in the Stark-Jacobi operator (\ref{Stark-Jacobi}) is 
equal to
\be \label{a1}a = \langle n|\sum_{x\in\Z} S^1_x|n-1\rangle = 
\frac{1}{2}\sum_{k\ge0} (-1)^k q^{k(k+1)} \frac{1+q^{1+2k}}{1-q^{1+2k}}\,.
\ee
\end{enumerate}
\end{lemma}

\begin{proof} Here, we only show the key steps for Part 2 and 3; the details 
can be found in \cite{St2}. The exponential localization of the measures 
$p(m)$ and $\tilde{p}(m)$ can be deduced from the explicit formulas but it 
can also be done much easier by using the following estimate
\be
\label{Bruno's est}
\langle{\psi_0}|[\mfr{1}{2} + \textrm{sgn}(x) S_x^3]|{\psi_0}\rangle
  \leq C q^{|x|}
\ee
for some $C=C(q)$ independent of $x$.
Let $\Omega^{+-} = \bigotimes_{x\le0} \binom{1}{0} \otimes \bigotimes_{x>0}
\binom{0}{1}$. We define the so-called grand-canonical states
\be \psi(z) = \prod_{x=-\infty}^0 (1+z^{-1} q^{-x} S^-_x)
              \prod_{x=1}^\infty (1+z q^{x} S^+_x) \Omega^{+-}.
\ee	      
The reason for this definition is that 
\be\label{grand canon}\psi(z) = \sum_{n\in \mathbb Z} \psi_n z^n .
\ee
It is clear that in the grand canonical ground state, for real $z=q^{-\mu}$,
$$
\frac{\langle{\psi(z)|S_x^3|\psi(z)}\rangle}{\langle{\psi(z)|\psi(z)}\rangle}
=\frac{1}{2} \cdot \frac{q^{(x-\mu)/2}-q^{(\mu-x)/2}}{\sqrt{q^{x-\mu}+
q^{\mu-x}}}\, ,
$$
which proves exponential localization of the interface in the grand 
canonical ground state. E.g., setting $\mu=0$, one has for $x>0$,
\be
\label{BrunoEstmtHelper}
\begin{split}
-\frac{1}{2} \leq \frac{1}{2} \cdot 
\frac{q^{x/2}-q^{-x/2}}{\sqrt{q^{x}+q^{-x}}}
&= -\frac{1}{2} + \frac{1}{2} \cdot 
\frac{\sqrt{1+q^{2x}}-1+q^x}{\sqrt{1+q^{2x}}}\\
  &\leq -\frac{1}{2} + \frac{1}{2} [1+\frac{1}{2} q^{2x} -1 + q^x]\\
  &\leq -\frac{1}{2} + \frac{3}{4} q^x\, ,
\end{split}
\ee
and a similar inequality holds for $x<0$.
Setting $\mu=0$, (i.e. $z=1$) we see that
$$
\langle{\psi(1)}|(\mfr{1}{2} \pm S_x^3)|{\psi(1)}\rangle
  = \sum_{n\in \mathbb Z} \langle{\psi_n}|(\mfr{1}{2} \pm S_x^3)|{\psi_n}
  \rangle
  \geq \langle{\psi_0}|(\mfr{1}{2} \pm S_x^3)|{\psi_0}\rangle\, ,
$$
using the fact that $(\mfr{1}{2} \pm S_x^3)$ is a nonnegative operator.
Hence, 
$$
\frac{\langle{\psi_0|[\mfr{1}{2} \pm S_x^3]|\psi_0}\rangle}{\langle{
\psi_0|\psi_0}\rangle}
\leq \frac{\|\psi(z)\|^2}{\|\psi_0\|^2} \cdot 
\frac{\langle{\psi(1)|[\mfr{1}{2} \pm S_x^3]|\psi(1)}\rangle}{
\langle{\psi(1)|\psi(1)}\rangle}\, .
$$
Since $(\|\psi(z)\|^2/\|\psi_0\|^2)<\infty$, and using equation 
(\ref{BrunoEstmtHelper}) we derive equation (\ref{Bruno's est}).
This proves Part 1.

Next, using the $q$-binomial product theorem we get that
\beax\langle\psi(z)|S_0^+|\psi(z)\rangle &=& z^{-1} \prod_{x=-\infty}^{-1}
(1+z^{-2}q^{-2x})\prod_{x=1}^{\infty}(1+z^{2}q^{2x}) 
\\
&=&z^{-1}\left(\sum_{n=0}^\infty |z|^{-2n} \frac{q^{n(n+1)}}{(q^2;q^2)_n} 
\right)\left(\sum_{n=0}^\infty |z|^{2n} \frac{q^{n(n+1)}}{(q^2;q^2)_n} 
\right)
\\
&=&z^{-1}\sum_{n=-\infty}^\infty |z|^{2n} \frac{q^{|n|(|n|+1)}
   f(q^{2(|n|+1)})}{(q^2;q^2)_\infty} \,.
\eeax
Comparing powers of $z$ in (\ref{grand canon}) we have
\beax\langle n|S_0^+|n-1\rangle
&=&\frac{\langle \psi_n|S_0^+|\psi_{n-1}\rangle}{\|\psi_{n-1}\|\|\psi_{n}\|} 
\\
&=& q^{|n|} f(q^{2|n|+2}) \,.
\eeax
We have used that $\|\psi_{n}\|^2 = q^{n(n+1)}/(q^2;q^2)_\infty$. Finally, 
\beax\sum_{x\in\mathbb Z}\langle n|S_x^+|n-1\rangle
&=&\sum_{x\in\mathbb Z}\langle x|S_0^+|x-1\rangle 
\\
&=&\sum_{x\in\mathbb Z} q^{|x|} \sum_{k\ge0} (-1)^k q^{2(|x|+1)k + k(k-1)}
\\
&=&\sum_{k\ge0} (-1)^k q^{k(k+1)} \frac{1+q^{1+2k}}{1-q^{1+2k}}\,.
\eeax
\end{proof}

\section{Stark-Jacobi Operator on $\mathbb Z^d$}

\begin{lemma} Let $Kf(\vec{n}) = (\a\D + \vec{\g}\cdot \vec{n}) f(\vec{n})$ 
be selfadjointly defined on ${\mathcal D}(K) = \{ f\in\ell(\mathbb Z^d) : 
(\vec{\g}\cdot \vec{n}) f(\vec{n}) \in\ell^2(\mathbb Z^d)\}$. If all 
$\g_j$ are non-zero and there is some $\g_0$ such that $\g_j = a_j \g_0$ with
$a_j \in \mathbb{Z}$ for all $j=1,\dots,d$, i.e.\ if $(\g_1,\dots,\g_d)$ are 
commensurable, then
$$
\textrm{spec}(K) = \frac{\|\g\|^2}{\lcm(a_1,\dots,a_d) \g_0} \mathbb Z\, ,
$$
where $\lcm(a_1,\dots,a_d)$ is the least common multiple.
If all $g_j$ are nonzero, but they are not commensurable, i.e.,
there exists a pair $\g_j$, $\g_k$ such that $\g_j/\g_k$ is irrational, 
then $\textrm{spec}(K) = \mathbb{R}$ and the spectrum is dense pure point 
spectrum. In this case there is an eigenvalue at every point of the lattice
$$
\sum_{j=1}^d \frac{\|\g\|^2}{\g_j} \mathbb{Z}
  = \{\sum_{j=1}^d a_j \|\g\|^2/\g_j : a_1,\dots,a_d \in \mathbb{Z}\}\, .
$$ 
However, if there are $1\le l<d$ non-zero components of 
$\vec{\g}$, say the first $l$, then 
$$
\textrm{spec}(K) = \cl\Big([-2\a(d-l),2\a(d-l)] + 
\sum_{j=1}^l (\frac{\|\g\|^2}{\g_j}\mathbb Z)\Big)\, ,
$$
where $\cl(.)$ is norm closure in $\R$,
and there is essential spectrum even when all $\gamma_j$ are commensurable.
\end{lemma}

\begin{proof} Let us assume first that all components are non-zero. 
In Fourier space we have the eigenvalue equation
$$\left(2\a \sum_{j=1}^d \cos{(k_j)} + i\|\g\| \sum_{j=1}^d 
\frac{\g_j}{\|\g\|} \frac{\p}{\p k_j} -\l\right) \phi(k) = 0\,.
$$  
Let $R=R(\g)$ be an orthogonal matrix such that $\sum_{j=1}^d R_{ij}
\frac{\g_j}{\|\g\|}=\d(i,1)$. (In particular this means 
$R_{1j} = \mfr{\gamma_j}{\|\gamma\|}$.) Setting $\tilde{k}=R k$, we get
$$\frac{\p}{\p\tilde{k}_1} \phi(\tilde{k}) = -\frac{i}{\|\g\|} 
\left(\l - 2\a \sum_{j=1}^d
\cos{\Big(\sum_{m=1}^d R_{mj}\tilde{k}_m\Big)}\right)\phi(\tilde{k})\,.
$$  
The solution is thus
$$\phi(\tilde{k}) = C e^{-\frac{i}{\|\g\|} \left(\l \tilde{k}_1 - 2\a 
\sum_{j=1}^d (R_{1j})^{-1} \sin{(\sum_{m=1}^d R_{mj}\tilde{k}_m)}\right)} \,.
$$
Using that $R_{1j}= \frac{\g_j}{\|\g\|}$ we have 
\be \phi(k) = C e^{-\frac{i}{\|\g\|} \left(\l \sum_{j=1}^d 
\frac{\g_j}{\|\g\|} k_j - 
2\a \sum_{j=1}^d \frac{\|\g\|}{\g_j} \sin{(k_j)}\right)} \,.
\ee		   
Finally, the periodicity condition, $\phi(k_j+2\pi) =\phi(k_j), 
j=1,\ldots,d$ requires that
$$\l \in \frac{\|\g\|^2}{\g_j}\mathbb Z \,.
$$

Now suppose that the first $l$ components of $\vec{\g}$ are non-zero and 
the rest are zero. According to the decomposition $\ell^2(\Z^d)=\ell^2
(\Z^l)\bigoplus\ell^2(\Z^{d-l})$, we write $K$ as the sum of Stark-Jacobi 
operator $K^{(l)}$ acting on $\ell^2(\Z^l)$ and 
$\a\D^{(l)}$ acting on $\ell^2(\Z^{d-l})$. Hence the result follows.
\end{proof}

\end{appendix}


\begin{thebibliography}{99}

\bibitem{AS}
M.~Abramowitz and I.A. Stegun, \emph{Handbook of mathematical functions}, Dover
  Books in Mathematics, 1970.

\bibitem{ASW}
F.~C. Alcaraz, S.~R. Salinas, and W.~F. Wreszinski, \emph{Anisotropic
  ferromagnetic quantum domains}, Phys. Rev. Lett \textbf{75} (1995), 930--933.

\bibitem{ARRS}
T.~Antal, Z.~R\'acz, A.~R\'akos, and G.M. Sch{\"u}tz, \emph{Transport in the
  {XX} chain at zero temperature: Emergence of flat magnetization profiles},
  Phys. Rev. E, \textbf{59} (1999), 4912--4919.

\bibitem{Araki2}
H.~Araki, \emph{On the {XY} model on two-sided infinite chain}, Publ. RIMS
  Kyoto Univ. \textbf{20} (1984), 277--296.

\bibitem{AP}
W.H. Aschbacher and C.-A. Pillet, \emph{Non-equilibrium steady states of the
  {XY} model}, Preprint, 2002, \mbox{mp-02-277}.

\bibitem{BG}
D.~Babbitt and E.~Gutkin, \emph{The {P}lancherel formula for the infinite {XXZ}
  {H}eisenberg spin chain}, Lett. Math. Phys. \textbf{20} (1990), no.~2,
  91--99.

\bibitem{BT}
D.~Babbitt and L.~Thomas, \emph{Ground state representation of the infinite
  one-dimensional {H}eisenberg ferromagnet. {II:} {A}n explicit {P}lancherel
  formula}, Commun. Math. Phys. \textbf{54} (1977), no.~3, 255--278.

\bibitem{CapMart2}
P.~Caputo and F.~Martinelli, \emph{Relaxation time of anisotropic simple
  exclusion processes and quantum {H}eisenberg models}, preprint, 2002,
  \mbox{math. PR/0202025}.

\bibitem{CNS}
P.~Contucci, B.~Nachtergaele, and W.L. Spitzer, \emph{The ferromagnetic
  {H}eisenberg {XXZ} chain in a pinning field}, Phys. Rev. B \textbf{66}
  (2002), Art.\# 064429.

\bibitem{GW}
C.-T. Gottstein and R.~F. Werner, \emph{Ground states of the infinite
  q-deformed {Heisenberg} ferromagnet}, Preprint, \mbox{cond-mat/9501123}.

\bibitem{KS}
R.~Killip and B.~Simon, \emph{Sum rules for Jacobi matrices and their applications 
to spectral theory}, \mbox{mp-arc/01-453}, to appear in Ann.~ of Math.

\bibitem{KN1}
T.~Koma and B.~Nachtergaele, \emph{The spectral gap of the ferromagnetic {XXZ}
  chain}, Lett. Math. Phys. \textbf{40} (1997), 1--16.

\bibitem{KN3}
\bysame, \emph{The complete set of ground states of the ferromagnetic {XXZ}
  chains}, Adv. Theor. Math. Phys. \textbf{2} (1998), 533--558,
  \mbox{cond-mat/9709208}.

\bibitem{KNS}
T.~Koma, B.~Nachtergaele, and S.~Starr, \emph{The spectral gap for the
  ferromagnetic spin-j {XXZ} chain}, Adv. Theor. Math. Phys. \textbf{5} (2001),
  1047--1090.

\bibitem{LSM}
E.~Lieb, T.~Schultz, and D.~Mattis, \emph{Two soluble models of an
  antiferromagnetic chain}, Ann. Phys. \textbf{16} (1961), 407--466.

\bibitem{Mat}
T.~Matsui, \emph{On ground states of the one-dimensional ferromagnetic {$XXZ$}
  model}, Lett. Math. Phys. \textbf{37} (1996), 397.

\bibitem{MPS}
M.~Moser, A.~Prets, and W.~Spitzer, \emph{Time evolution of spin waves}, Phys.
  Rev. Lett. \textbf{83} (1999), 3542--3545.

\bibitem{Oga}
Y.~Ogata, \emph{The diffusion of the magnetization profile in the {XX}-model},
  Preprint, 2002, \mbox{cond-mat/0210011}.

\bibitem{SS}
R.~H. Schonmann and S.~Shlosman, \emph{Wulff droplets and the metastable
  relaxation of kinetic {Ising} models}, Commun. Math. Phys. \textbf{194}
  (1998), 389--462.

\bibitem{St2}
S.~Starr, \emph{Some properties of the low lying spectrum of the quantum {XXZ}
  spin system}, Ph.D. thesis, U.C. Davis, Davis, CA 95616, June 2001,
  \mbox{math-ph/0106024}.

\end{thebibliography}
\providecommand{\bysame}{\leavevmode\hbox to3em{\hrulefill}\thinspace}

\end{document}